\documentclass[12pt]{article}
\usepackage{paper2e}
\usepackage{mydefs2e}
\usepackage{epsf}

\newcommand{\dsb}{dynamical \susy breaking\xspace}

\begin{document}
\begin{titlepage}

\preprint{UMD-PP-97-115\\
{hep-ph/9706235}}

\title{Na\"\i ve Dimensional Analysis\\\medskip
and Supersymmetry}

\author{Markus A. Luty}

\address{Department of Physics\\
University of Maryland\\
College Park, Maryland 20742}

\begin{abstract}
In strongly-coupled theories with no small parameters,
there are factors of $4\pi$ that appear when the couplings
of the low-energy effective \lag are written in units of the
effective cutoff $\La$.
These numerical factors can be explained using ``\naive
dimensional analysis.''
We extend these ideas to \susc theories,
and show how to systematically include small parameters and
couplings to weakly-interacting fields.
The basic principle is that if the fundamental theory is
strongly coupled, then the effective theory must also be strongly
coupled at the scale $\La$.
We use our results to analyze several examples where strong \susc
dynamics may be relevant for phenomenology.
For models that break \susy through strong dynamics with
no small parameters, we show that the Goldstino decay constant
$F$ is given by $F \sim \La^{2} / (4\pi)$.
We also consider theories with standard-model gauge
bosons coupled directly to strong \susy-breaking dynamics
near the weak scale;
smoothly-confining theories;
and a model that breaks \susy through the mechanism of a
deformed moduli space.
%
%
\end{abstract}

\date{June, 1997}

\end{titlepage}

\section{Introduction}
If nature is \susc at short distances, then \dsb provides an attractive
possible explanation for the origin of the \susy breaking scale.
Many interesting models of \dsb involve non-calculable strong
dynamics, and one would like to construct the 
low-energy effective theory for such models and estimate the couplings
involved.
If there are no small parameters in the theory and the only scale
is the energy where the dynamics becomes strong, then one might
expect all  couplings in the low-energy theory to be order one in
units of the scale of strong dynamics.
However, our experience with QCD suggests that the situation is not 
so simple.
In QCD, there are differences by factors of order 10 in quantities
that are formally of order $\La_{\rm QCD}$.
For example, the pion decay constant is $f \sim 100\MeV$, while the scale
that controls the convergence of chiral perturbation theory is of order
$1\GeV$.

These numerical factors can be counted using
``na\"\i ve dimensional analysis'' (NDA) \cite{NDA}.
The starting point of NDA is to assume that there is a single
mass scale $\La$ in the fundamental theory that sets the
energy scale for the states that are integrated out to obtain
the effective theory.
One then assumes that subleading corrections will be suppressed
compared to leading terms by powers of $E / \La$, 
where $E$ is the energy scale of the observable of interest.
The factors of $4\pi$ that appear can then be counted using
the renormalization properties of the effective
theory \cite{Weinberg} and naturalness arguments.
For example, if the strong dynamics breaks a global
symmetry, the decay constant of the Nambu--Goldstone bosons is
given by $f \sim \La / (4\pi)$.
This counting accounts for some striking facts about QCD:
for example, the existence of an approximate $\SU{3}$ flavor
symmetry despite the fact that (numerically)
$m_{s} \sim 100\MeV \sim f$.

In this paper, we first apply NDA to theories that spontaneously break
\susy through strong dynamics at a scale $\La$.
(Examples of such theories are $\SU{5}$ gauge theory with a $10$ and a
$\bar{5}$ \cite{ADSsu},
or $\SO{10}$ gauge theory with a single
spinor \cite{ADSso}.)
We argue that the Goldstino decay constant $F$ (with dimension 
mass-squared) is given by $F \sim \La^2 / (4\pi)$.
We then extend this counting to include 
weakly-coupled fields and small parameters, and also consider
theories with confining dynamics.

In all cases, the principle we use is that if the fundamental
theory is strongly coupled at the scale $\La$, then the effective
theory must also be strongly coupled at this scale.
We assume that strong coupling means that all loops are equally
important (``loop democracy'').
Even if we cannot perform the matching calculation at the scale $\La$,
we can count the factors of $4\pi$ that must appear in the effective
theory so that the dynamics is strong at the matching scale.

This paper is organized as follows.
In Section 2, we briefly review the arguments for NDA in 
nonsupersymmetric theories with spontaneously broken symmetries.
In Section 3, we extend these arguments to spontaneous breaking of 
\susy.
In Section 4, we show how to include small parameters and
couplings to weakly-interacting fields.
In Section 5, we analyze some examples where the $4\pi$ counting
discussed above may be relevant for phenomenology and model-building.
We first consider theories with standard-model gauge
bosons coupled directly to strong \susy-breaking dynamics
near the TeV scale.
We then consider models with confining dynamics \cite{confine},
and analyze a model that breaks \susy through
the mechanism of a deformed moduli space \cite{deform}.
Section 6 contains our conclusions.

\section{Power Counting for Nambu--Goldstone Bosons}
\subsection{Linear Sigma Model}
We begin our review of NDA for global symmetry breaking with the
linear sigma model.
We consider a theory of a complex scalar field $\Phi$ with \lag
\beq[linsig]
\scr{L} = \partial^{\mu} \Phi^{\dagger} \partial_{\mu} \Phi
- \sfrac{1}{2} \la (\Phi^{\dagger} \Phi - v^{2})^{2}.
\eeq
If $v^{2} > 0$, then $\avg{\Phi} = v$ and the $\U1$ global symmetry
is spontaneously broken.
The theory contains a massless Nambu--Goldstone boson
with decay constant $f = v$, and a 
massive scalar (``$\si$'') with mass $m_{\si}^{2} = \la v^{2}$.
We therefore identify $\La \sim m_{\si}$ as the scale 
of ``new physics'' in the low-energy effective \lag.

It is straightforward to integrate out the heavy scalar field at tree
level by making the field redefinition
\beq
\Phi = v \hat{\si} e^{i \hat{\Pi}}
\eeq
and using the $\hat{\si}$ equations of motion.
To order $\partial^{4}$, we obtain the effective \lag
\beq[sigcount]
\scr{L}_{\rm eff} = v^{2} \hat{\scr{L}}_{\rm eff},
\qquad
\hat{\scr{L}}_{\rm eff} = 
\partial^{\mu} \hat{\Pi} \partial_{\mu} \hat{\Pi}
+ \frac{1}{2 m_{\si}^{2}}
\left( \partial^{\mu} \hat{\Pi} \partial_{\mu} \hat{\Pi} \right)^{2}
+ O(\partial^{6}).
\eeq
To this order, $v^{2}$ appears as an overall factor in the effective
lagrangian, and all couplings in $\hat{\scr{L}}_{\rm eff}$ are order
1 in units where $m_{\si} = 1$.

In fact, the scaling in \Eq{sigcount}
holds to all orders in the derivative expansion.
This is because the fundamental \lag \Eq{linsig} can be written
\beq[sigscale]
\scr{L} = v^{2} \hat{\scr{L}},
\qquad
\hat{\scr{L}} =
\partial^{\mu} \hat{\si} \partial_{\mu} \hat{\si}
+ \hat{\si}^{2} \partial^{\mu} \hat{\Pi} \partial_{\mu} \hat{\Pi}
- \frac{m_{\si}^{2}}{2} (\hat{\si}^{2} - 1)^{2}.
\eeq
Again, $v^{2}$ appears as an overall factor and all couplings in
$\hat{\scr{L}}_{\rm eff}$ are  order 1 in units where
$m_{\si} = 1$.
Since the overall factor does not affect the classical equations of 
motion, it is easy to see that integrating out the field $\hat{\si}$
at tree level results in an effective \lag of the form \Eq{sigcount}
to all orders in the derivative expansion.
This shows that the effective \lag has the form
\beq[sigNDA]
\scr{L}_{\rm eff} \sim v^{2}
\sum_{p,q} \left( \frac{\partial_{\mu}}{m_{\si}} \right)^{p}
\left( \vphantom{\frac{\partial_{\mu}}{m_{\si}}}
\frac{\Pi}{v} \right)^{q},
\eeq
where we have written the result in terms of the canonically normalized
field $\Pi \equiv v \hat{\Pi}$.
\Eq{sigNDA} is the NDA scaling for the effective 
\lag (with the identifications $f \sim v$,
$\La \sim m_{\si}$) \cite{NDA}.

The reason that the effective \lag exhibits the scaling
\Eq{sigNDA} is that there are only two scales in the full theory.
In a more complicated sigma model, we could have several widely
different scales, invalidating the power-counting formula \Eq{sigNDA}.
However, if the parameters of the model are adjusted so that all 
nonzero tree masses are of the same order, then 
arguments such as these show that 
the NDA scaling is also valid in these more general theories.
(We will give the analogous arguments in detail for the case of
\susy breaking below.)
We therefore expect the NDA scaling to hold in all effective theories
with a single scale of ``new physics.''

\subsection{Strong Coupling}
We can try to get some insight into strongly-coupled theories from
the linear sigma model above by allowing the coupling $\la$ to become
large.
When $\la$ (defined at a suitable subtraction scale) becomes sufficiently
large, the perturbative Landau pole of the theory is near the
physical mass of the $\si$ particle.
It is unlikely that this theory
can be made sensible as an effective theory at the scale $m_\si$.
Nonetheless, we can make sense out of this theory by imposing a cutoff
of order $m_\si$ on the theory, and study the dynamics below the scale
$m_\si$.
This procedure is certainly far from rigorous, but in this way
we can hope to get some qualitative insight into the low-energy
effective theory resulting from strong dynamics.

Because $v^{2}$ appears only as an overall factor in \Eq{sigscale},
$1/v^2$ acts as a loop counting parameter analogous to $\hbar$.
The remaining dimensions are made up by $m_\si$ and powers of momentum,
so loop effects in the fundamental theory at the scale $m_\si$ are
suppressed by powers of
\beq
\ell = \frac{1}{16\pi^{2}} \frac{m_{\si}^{2}}{v^{2}}
\sim \frac{\la}{16\pi^{2}}
\eeq
compared to tree effects.
This theory becomes strongly coupled when all loop effects are equally
important (``loop democracy''), which means $\ell \sim 1$.
Note that in this limit, we have $m_{\si} \sim 4 \pi v$,
which corresponds to the usual NDA estimate $\La \sim 4\pi f$ for
strong theories with no small parameters.

Even though the dynamics at the scale $\La$ becomes
non-calculable for $\La \sim 4\pi f$, the effective theory given by
\Eq{sigNDA} is still predictive at scales small compared to $\La$.
The reason is that in the effective theory the Nambu--Goldstone
bosons are
derivatively coupled and $m_\si$ does not appear
as a kinematic scale in the loop diagrams.
Loops in the effective theory are therefore
suppressed by powers of
\beq
\ell_{\rm eff} = \frac{p^2}{16\pi^2 v^2},
\eeq
where $p$ is an external momentum.
Loop effects are small for
$p \ll \La$, and we have a predictive low-energy expansion.
This reasoning embodies an appealing picture of the origin of the
factors of $4\pi$:
they are put in the effective lagrangian so that the effective
description is strongly coupled at the scale $\La$ where it matches
onto the fundamental theory.
The effective theory is weakly coupled at energies below $\La$ because
the Nambu--Goldstone bosons are derivatively coupled.

One might wonder whether it is possible to have $\La \gg 4\pi f$
in the effective theory.
This is unnatural because the loop
corrections logarithmically renormalize higher-derivative coefficients
in the effective \lag.
Therefore the condition $\La \gg 4\pi f$ could hold only
at a specific value of the renormalization scale, which is clearly
unphysical \cite{NDA}.

The loop-counting parameter in the effective theory may be
smaller than the estimates above in a strongly-coupled theory if there
is a small parameter in the theory.
For example, in QCD for a large number of colors $N$,
the loop counting parameter in the low-energy theory is
$1 / N \ll 1$.
In this case, the relation between the decay constant and the scale
of new physics is $\La \sim 4\pi f / \sqrt{N}$ \cite{NNDA}.


\section{Power Counting for Goldstinos}
\subsection{A Simple O'Raifeartaigh Model}
We begin our discussion of \susy breaking with a model containing
chiral superfields $\Phi$, $S$, and $\Si$, with superpotential
\beq
W = \ka S + \sfrac{1}{2} \la S \Phi^{2} + M \Si \Phi.
\eeq
We choose all couplings to be real and positive by rescaling the 
fields.

\Susy is spontaneously broken by $\avg{F_S} = \ka$.
$\avg{S}$ is undetermined at tree level, corresponding to the 
tree-level flat direction that exists in all \OR models.
The global minimum of the tree potential is at
$\avg{\Phi} = \avg{\Si} = 0$ provided that
$M^{2} \ge \la \ka$.
This is the parameter range we will consider.

The spectrum of this theory is as follows.
The fermionic components of $\Phi$ and $\Si$ get a tree-level
Dirac mass $M$, and the $S$ fermion is exactly massless
(it is the Goldstino).
The complex $\Si$ scalar gets a tree-level mass $M$,
and the two real $\Phi$ scalars get tree-level
mass-squared $M^{2} \pm \la\ka$.
(The splitting $\la\ka$ in the scalar masses breaks \susy
in the tree-level spectrum.)
The $S$ scalar is massless at tree level, and gets a positive
mass-squared at one loop of order
\beq[Smass]
m^{2}_{S} \sim \frac{\la^{2}}{16\pi^{2}} \la\ka.
\eeq
(The global minimum is at $\avg{S} = 0$.)
For weak coupling, $m_{S}$ is much smaller than the other masses in 
the problem, and we do not obtain a simple scaling
analogous to the NDA result for global symmetry breaking.
However, we will argue that a simple scaling emerges in a
strong-coupling limit where all heavy particles have comparable
mass.

We first choose the parameters so that all the non-zero
tree-level masses are of the same order.
This means
\beq[lacond]
M^{2} \sim \la \ka.
\eeq
We then define rescaled component fields and couplings
\beq
\hat{\phi} \equiv \frac{M}{\ka} \phi,
\qquad
\hat{\psi} \equiv \frac{M^{1/2}}{\ka} \psi,
\qquad
\hat{\la} \equiv \frac{\ka}{M^{2}} \la,
\eeq
and write the \lag as
\beq[ORrescale]
\bal
\scr{L} = \ka^{2} \biggl\{&
\frac{i}{M} \hat{\psi}_{\Phi}^{\dagger} \bar{\si}^{\mu}
\partial_{\mu} \hat{\psi}_{\Phi}
+ \frac{1}{M^{2}} \partial^{\mu} \hat{\phi}_{\Phi}^{\dagger}
\partial_{\mu} \hat{\phi}_{\Phi}
+ \cdots
\\
& - ( \hat{\psi}_{\Phi} \hat{\psi}_{\Si} + \hc )
- \hat{\la} (
\sfrac{1}{2} \hat{\phi}_{S} \hat{\psi}_{\Phi} \hat{\psi}_{\Phi}
+ \hat{\phi}_{\Phi} \hat{\psi}_{\Phi} \hat{\psi}_{S} + \hc )
\\
&- | \sfrac{1}{2} \hat{\la} \hat{\phi}_{\Phi}^{2} - 1 |^{2}
- | \hat{\phi}_{\Phi} |^{2}
- | \hat{\la} \hat{\phi}_{S} \hat{\phi}_{\Phi} + \hat{\phi}_{\Si} |^{2}
\biggr\}.
\eal\eeq
The condition that the tree masses are of the same order 
means that $\hat{\la} \sim 1$ (see \Eq{lacond}), so the \lag 
has the form
\beq
\scr{L} = \ka^{2} \hat{\scr L},
\eeq
where all the couplings in $\hat{\scr L}$ are order 1 in units where
$M = 1$.

We see that the parameter $1/\ka^{2}$ acts as a loop counting parameter 
if we work in units where $M = 1$, so loops are suppressed by powers of
\beq
\ell = \frac{1}{16\pi^{2}} \frac{{M}^{4}}{\ka^{2}}
\sim \frac{\la^{2}}{16\pi^{2}}.
\eeq
We now consider the na\"\i ve strong-coupling limit $\ell \sim 1$.
We then find that the $S$ mass is of order $M$ (see \Eq{Smass}),
the same as the other heavy masses.
Integrating out the heavy particles gives rise to an 
effective \lag for the Goldstino of the form
\beq[normLa]
\scr{L}_{\rm eff} = \ka^{2} \hat{\scr L}_{\rm eff},
\eeq
where all couplings in $\hat{\scr L}_{\rm eff}$ are order 1 in units of 
$M$.
This means
\beq[normLaToo]
\hat{\scr L}_{\rm eff} \sim \sum_{p,q}
\left( \frac{\partial_{\mu}}{M} \right)^{p}
\left( \hat{\chi} \right)^{q},
\eeq
where $\hat{\chi}$ is the (rescaled) Goldstino field.
Defining a canonically-normalized Goldstino field
$\chi \equiv \ka \hat{\chi} / M^{1/2}$ 
and identifying $F \sim \ka$ and $\La \sim $M,
the effective \lag can be written
\beq[SUSYNDAform]
\scr{L}_{\rm eff} \sim F^{2} \sum_{p,q}
\left( \frac{\partial_{\mu}}{\La} \right)^{p}
\left( \frac{\La^{1/2} \chi}{F} \right)^{q},
\eeq
with $\La^2 \sim 4\pi F$.
This is the NDA form of the Goldstino \lag.
It is easy to see that this scaling is compatible with the constraints
on the couplings coming from the non-linear realization of \susy
\cite{nlSUSY}.

\subsection{General O'Raifeartaigh Model}
We now extend this analysis to arbitrary \OR models.
We assume the theory has a canonical kinetic term%
\footnote{The argument is easily extended to theories with non-trivial
\Kahler potentials.}
\beq
\scr{L}_{D} = \myint d^2\th d^2\bar{\th}\,
\Phi^{\dagger}_{a} \Phi^{a},
\eeq
and F terms
\beq
\scr{L}_{F} = \myint d^{2}\th\,
\sum_{n = 1}^{\infty} \frac{1}{n!} \la^{(n)}_{a_{1} \cdots a_{n}}
\Phi^{a_{1}} \cdots \Phi^{a_{n}} + \hc
\eeq
We will show that the couplings can be chosen so that all
tree-level masses are of the same order $\La$, and derive the scaling
of the low-energy theory.
We will work in units where $\La = 1$.
We begin by defining rescaled superfields
\beq[ORfieldrescale]
\Phi \equiv F \hat{\Phi},
\eeq
in terms of which the kinetic term is
\beq
\scr{L}_{D} = F^{2} \myint d^2\th d^2\bar{\th}\,
\hat{\Phi}^{\dagger}_{a} \hat{\Phi}^{a}.
\eeq
In order to make all physical masses of order 1, we write the 
superpotential couplings
\beq
\la^{(n)}_{a_{1} \cdots a_{n}} = F^{2 - n}
\hat{\la}^{{(n)}}_{a_{1} \cdots a_{n}},
\eeq
and choose values%
\footnote{More precisely, what is meant here is that all 
\emph{nonzero} couplings are of order 1.
If an \OR model breaks \susy, it will break it for all values the
parameters except for isolated critical values where the limiting
behavior of the potential for large field values changes \cite{witten}.
We can therefore always find a \susy-breaking vacuum where the 
non-zero couplings $\hat{\la}^{(n)}$ are order 1.}
\beq[lamparam]
\hat{\la}^{(n)} \sim 1.
\eeq
The full \lag then has the form
\beq[genORscale]
\scr{L}_{D} + \scr{L}_{F} = F^{2} \hat{\scr{L}},
\eeq
where all couplings in $\hat{\scr{L}}$ are of order 1.
Since $F$ scales out of the classical equations of motion, the 
tree-level masses are independent of $F$, and hence the non-zero 
tree \vevs and masses are all of order 1.%
\footnote{In ``inverse hierarchy'' models \cite{invert}, there are
mass hierarchies of order $e^{-16\pi^2 / \la^2}$, where $\la$ is
a superpotential coupling.
These hierarchies disappear in the strong-coupling limit considered
below, and we 
expect our final results to hold even in this class of models.}
With the choice of parameters \Eq{lamparam},
the \susy breaking order parameter is
\beq
\left\langle \frac{\partial W}{\partial\Phi^{a}} \right\rangle
= F \sum_{n = 1}^{\infty}
\frac{1}{(n - 1)!} \hat{\la}^{(n)}_{a b_{2} \cdots b_{n}}
\langle \hat{\Phi}^{b_{2}} \cdots \hat{\Phi}^{b_{n}} \rangle
\sim F.
\eeq

The loop-counting parameter is
\beq
\ell \sim \frac{1}{16\pi^{2} F^{2}}.
\eeq
For $\ell \ll 1$, the theory has a light complex scalar with mass
of order $\ell$, corresponding 
to the tree-level flat direction that exists in all \OR models.
To get a theory where all massive particles have masses of the same 
order, we take the \naive strong coupling limit $\ell \sim 1$.
In this case, the effective \lag below the scale $\La$ 
contains only the Goldstino field, with the scaling
\beq
\scr{L}_{\rm eff} \sim F^{2} \hat{\scr{L}}_{\rm eff}.
\eeq
This leads to the NDA form \Eq{SUSYNDAform} for the effective 
\lag.

\subsection{Global Symmetry Breaking}
In the above analysis, note that the \vevs of the fields $\hat{\Phi}$ 
were order 1.
This means that
\beq
\avg{\Phi} \sim \frac{F}{\La} \sim \frac{\La}{4\pi}.
\eeq
Identifying $f \sim \avg{\Phi}$, this agrees with the estimate
$\La \sim 4\pi f$, so the power-counting above is consistent with the
NDA arguments for global symmetry breaking.

\section{Spurion Analysis}
\newcommand{\OF}{\scr{O}_{{F},d}}
\newcommand{\OFhat}{\hat{\scr{O}}_{{F},d}}
\newcommand{\JF}{\scr{J}_{{F},d}}
\newcommand{\JFhat}{\hat{\scr{J}}_{{F},d}}
\newcommand{\OD}{\scr{O}_{{D},d}}
\newcommand{\ODhat}{\hat{\scr{O}}_{{D},d}}
\newcommand{\JD}{\scr{J}_{{D},d}}
\newcommand{\JDhat}{\hat{\scr{J}}_{{D},d}}
We now extend these power-counting arguments to include small
coupling constants and couplings to weakly-interacting light fields.
We begin with the general \OR model discussed in the previous section.
We now add to the fundamental theory couplings of the form
\beq
\de\scr{L} = \myint d^{2}\th d^{2}\bar{\th}\, \JD \OD
+ \left[ \myint d^{2}\th\, \JF \OF + \hc \right],
\eeq
where $\OD, \OF \sim \Phi^d$ are dimension-$d$ operators composed
of strongly-interacting chiral superfields.
$\JD$ and $\JF$ may be coupling constants or products of coupling
constants and weakly-coupled superfields.
If $\JD$ or $\JF$ involve weakly-coupled superfields, the leading
contributions to the effective \lag arise from diagrams with no
loops involving weakly-interacting superfields.
In either case, $\JD$ ($\JF$) may be viewed as a source (or ``spurion'')
coupled to the operator $\OD$ ($\OF$).

We now write the operators $\OD$ and $\OF$ in terms of the rescaled fields
$\hat{\Phi}$ defined in \Eq{ORfieldrescale}
\beq
\ODhat \equiv F^{-d} \OD \sim \hat{\Phi}^d,
\quad
\OFhat \equiv F^{-d} \OF \sim \hat{\Phi}^d,
\eeq
and rescale $\JD$ and $\JF$ according to
\beq
\JDhat \equiv F^{d - 2} \JD,
\quad
\JFhat \equiv F^{d - 2} \JF.
\eeq
In terms of these, the coupling to the spurions can be written
\beq
\de \scr{L} = F^{2} \left[ 
\myint d^{2}\th d^{2}\bar{\th}\,
\JDhat \ODhat + 
\myint d^{2}\th\, \JFhat \OFhat
+ \hc \right].
\eeq
The point of this rescaling is that $\JDhat$ and $\JFhat$ now appear
in the fundamental \lag in the same way as a rescaled
strongly-coupled field $\hat{\Phi}$.
We therefore have (in units where $\La = 1$)
\beq
\bra{0} T \hat{\scr{O}}_{d_{1}} \cdots
\hat{\scr{O}}_{d_{n}} \ket{0}
\sim F^{2} + {\rm loop\ corrections}
\eeq
for any operators with dimensions $d_1, \ldots, d_n$.
This counting is reproduced in the effective \lag by
\beq
\scr{L}_{\rm eff} \sim F^{2} \sum_{p, \ldots, t}
\left( \frac{\partial_{\mu}}{\La} \right)^{p}
\left( \frac{\La^{1/2}\chi}{F} \right)^{q}
\left( \frac{1}{\La^{1/2}} \frac{\partial}{\partial\th} \right)^{r}
\left( \frac{F^{d - 2}}{\La^{d - 2}} \JD \right)^{s}
\left( \frac{F^{d - 2}}{\La^{d - 1}} \JF \right)^{t},
\eeq
where we have put back factors of $\La$ by dimensional analysis.
The derivatives with respect to the superspace coordinate
$\th$ are used to project out the component fields of the
spurions $\JD$ and $\JF$.

It is worth emphasizing that the arguments above are simply
a compact way of taking the \naive strong coupling limit of
\OR models.
The scaling above holds only if the spurion $\scr{J}$ in the
fundamental theory is coupled to
operator $\scr{O}$ that is a simple product of strong chiral
superfields.
It is straightforward to analyze general operators in the same way.
The result (in units where $\La = 1$) is that $\scr{J}$ appears in
the effective lagrangian with coefficient $F^{w - 2}$,
where $w$ is the sum of ``weights'' of the strong fields appearing in
$\scr{O}$.
Strong chiral fields and factors of the strong gauge field strength
$W_\al$ have weight 1, and gauge-covariant derivatives
(superspace or spacetime) have weight 0.

A counting of $4\pi$'s was given (without derivation) for
dynamically-generated $D$ terms in \Ref{effSUSY}, for
dimensionless couplings to weakly-coupled fields
($\JF$ with $d = 2$ in our notation).
We agree with the results of \Ref{effSUSY} in this case.

\section{Examples}
\subsection{Supersymmetry Breaking in the Observable Sector}
An interesting application of these ideas is to models of
gauge-mediated \susy breaking \cite{gmsb}.
In these models, \susy breaking is communicated to superpartners of 
observed fields by strong and electroweak gauge interactions.
Explicit models of this type usually involve a weakly-coupled
``messenger'' sector that separates the observable sector
from the \susy breaking dynamics.
These models have the virtue of avoiding many phenomenological problems
(although they may have problems with color breaking \cite{colorbreak}),
but they are rather complicated.
We will consider the possibility that the standard
model gauge fields couple directly to strong \susy-breaking dynamics
near the weak scale.%
\footnote{There are models in the literature in which the electroweak
gauge bosons are coupled directly to a \susy breaking sector
with a large scale hierarchy
\cite{PoppitzTrevedi}.
In these models, the scale of strong dynamics $\La$ is far above the
weak scale.}
This would require \susy-breaking gauge theory with an unbroken symmetry 
that is large enough to contain the standard-model gauge group.
On the other hand, the model should not have extra symmetries
that give rise to unwanted massless particles.
If one is interested in maintaining grand unification, the models 
must not contain too many fields carrying standard-model quantum numbers.
It may be difficult to construct explicit models satisfying all these
constraints, but we can easily analyze the low-energy theory resulting
from such a model using the ideas of this paper.
We will find that the phenomenology of these models is very similar 
to that of weakly-coupled messenger models, with $\La \sim 10\TeV$.

\epsfig{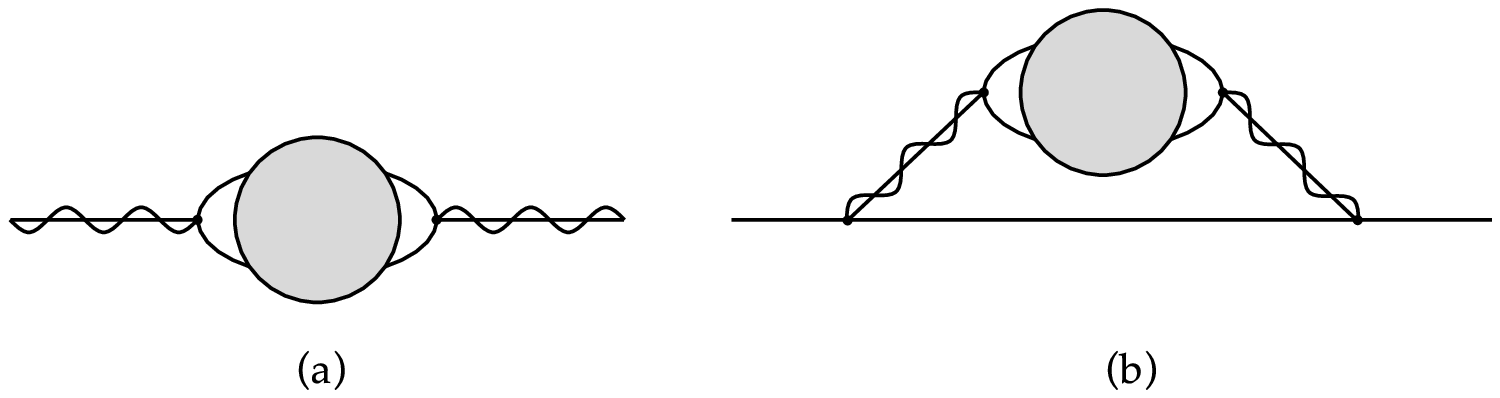}%
{Typical diagrams contributing to (a) gaugino masses and (b) scalar
masses in theories where fields are coupled weakly to the strongly
interacting fields.
The shaded blob denotes strong interactions.}

The term in the fundamental \lag that couples the standard-model
gauginos to the strongly-interacting fields is
\beq
\de\scr{L} = \myint d^{2}\th d^{2}\bar{\th}\,
\Phi^{\dagger} g V_{A} T_{A}^{(r)} \Phi,
\eeq
where $\Phi$ is a strongly-interacting field in the \rep $r$ of the 
standard-model gauge group,
$V_{A}$ is the weakly-interacting gauge superfield,
and $T_{A}^{(r)}$ is the gauge generator.
The gaugino mass arises from the diagram of Fig.~1a.
The operator $g V_A T_A^{(r)}$ acts as a source for a dimension-2
operator, so we have%
\footnote{It may appear surprising that the gaugino mass is proportional 
to $F^{2}$ rather than $F$, since the gaugino mass can be written as 
an F-term with a \susy-breaking spurion $\th\th$.
However, the powers of $F$ and $\La$ in our formulas serve only to count
powers of $4\pi$.
Also, our results are (by construction) consistent with the
strong-coupling limit of weakly-coupled models.}
\beq[stronggaugino]
m_{\la} \sim g^{2} I_{2}(r) \frac{F^{2}}{\La^{3}}
\sim \frac{g^{2} I_{2}(r)}{16\pi^{2}} \La,
\eeq
where $I_{2}(r)$ is the index of the \rep $r$.
Requiring that the weak gaugino masses be of order $100\GeV$ gives
$\La \sim 10\TeV$.

The scalar masses arise through the diagrams such as the one
in Fig.~1b.
This effect can be estimated as a 1-loop perturbative diagram with 
the insertion of non-perturbative gaugino mass and wavefunction 
renormalization.
This gives rise to a scalar mass-squared of order
\beq
m^{2}_{\phi} \sim I_{2}(r) C_{2}(r)
\left( \frac{g^{2}}{16\pi^{2}} \right)^{2} \La^{2},
\eeq
where $C_{2}(r)$ is the quadratic Casimir invariant of the \rep $r$.

These estimates are the same order of magnitude as the minimal 
messenger sector with singlet $S$ coupled
to messenger ``quarks'' $q$ and $\bar{q}$ 
via a superpotential term $\la S q \bar{q}$,
provided we make the identifications
\beq
F \sim \avg{F_S},
\quad
\La^2 \sim m^2_q \sim \la^2 \avg{S}^2 \sim \la \avg{F_S},
\quad
\la \sim 4\pi.
\eeq
The main difference between the strongly-coupled models considered
here and perturbative messenger models is that we expect
a rich spectrum of strong resonances near the scale
$\La \sim 10\TeV$ in the strong models.

%
%

\subsection{Confining Theories}
Another interesting class of \susc theories are those that confine and
have light composite chiral superfields \cite{confine}.
Models have been constructed where the composite particles have the
quantum numbers of quarks, leptons, and Higgs fields, and attempts
at model-building have been made
\cite{confinemodels,lastconfine}.
Confining models can break \susy through strong dynamics \cite{deform},
or they may be coupled to a separate \susy breaking sector in some way.

We begin by considering confining theories in which \susy is not broken.
The prototype model of this kind is \susc QCD with gauge group $\SU{2}$
and $6$ fields $Q$ in the fundamental representation.
Seiberg has argued convincingly that the low-energy effective theory
near the origin of moduli space of this model has a confined
description in terms of composite superfields with the quantum numbers
of the operators
$M^{jk} \sim \ep_{ab} Q^{a j} Q^{b j}$,
where $\ep_{ab}$ is the $\SU{2}$ metric and $j, k = 1, \ldots, 6$.
%
%
The effective theory has a dynamically-generated superpotential
\beq[cubicWeff]
W_{\rm eff} = \Pf(M)
\propto \ep_{j_1 \cdots j_6} M^{j_1 j_2} M^{j_3 j_4} M^{j_5 j_6}.
\eeq

This theory is controlled by strong dynamics with no small parameters,
so the arguments above lead us to expect that the effective
\lag should have the form
\beq[shouldform]
\scr{L}_{\rm eff} = \frac{1}{16\pi^2} \hat{\scr{L}}_{\rm eff},
\eeq
where all couplings in $\hat{\scr{L}}_{\rm eff}$ are order 1 in
units where $\La = 1$.%
\footnote{There are $\SU{N}$ and $\Sp{2N}$ generalizations of this
theory, and if $N \gg 1$ these theories have an additional small
parameter.
We will discuss this elsewhere.}
In terms of canonically normalized composite fields $M$, the
lagrangian has the form%
\footnote{The counting for the \Kahler potential
was given (without derivation) in \Ref{effSUSY}.}
\beq
\scr{L}_{\rm eff} = \frac{1}{16\pi^2} \left[
\myint d^2\th d^2\bar{\th}\,
K_{\rm eff} \left( \frac{4\pi M}{\La}, \frac{D_\al}{\La^{1/2}},
\ldots \right)
+ \myint d^2\th\, W_{\rm eff} \left( \frac{4\pi M}{\La},
\ldots \right)
\right].
\eeq
Couplings to other fields can be included using the results of the
previous section.

In any confining model, we therefore expect that dynamically generated
cubic couplings in the superpotential such
as the one in \Eq{cubicWeff} is of order $4\pi$ (not 1) when expressed
in terms of canonically normalized fields.
This means that confined description of this theory is not
weakly coupled at the scale $\La$.
However, we expect the cubic coupling to renormalize logarithmically
to zero in the infrared, so this theory is still weakly coupled at
energies far below $\La$.
If we want to identify a cubic term in a dynamically-generated
superpotential with the top-quark Yukawa coupling, the compositeness
scale must be large enough that the top-quark Yukawa coupling can
become perturbative at the weak scale.
The top quark mass is then controlled by an approximate
infrared fixed point, as in \Refs{BHL}.
(These points have also been noted in \Ref{lastconfine}, which appeared
as this paper was nearing completion.)

The confining theory described here is related to other theories
by adding mass terms, and we must check that our estimates
are consistent with these relations.
In fact, by adding mass terms for all but 2 $Q$'s, one obtains
a model whose non-perturbative dynamics far from the origin in
moduli space can be analyzed by a direct instanton calculation
\cite{adsinstant}.
However, to use this solution to find the physical couplings in the
confining theory near the origin of moduli space, it is necessary to
know the \Kahler potential close to the origin
to determine the normalization of the fields.
A similar ambiguity will affect any attempt to relate the physical
superpotential couplings to a weak-coupling calculation, and we
therefore believe that there is no conflict between our arguments
and the well-established relations between theories with different
numbers of colors and flavors.

\subsection{Dynamical Supersymmetry Breaking on a Deformed Moduli Space}
We now consider the model of \Refs{deform} as an
example where the factors of $4\pi$ are rather non-trivial.
The model has gauge group $\SU{2}$ with 4 fields $Q$ in 
the fundamental \rep.
There are also gauge singlet fields
$S_{jk} = -S_{kj}$ ($j = 1, \ldots, 4$), coupled to the $Q$'s via
the superpotential
\beq
W = \la S_{jk} \ep_{ab} Q^{aj} Q^{bk}.
\eeq
If we set $\la = 0$, Seiberg has argued that this theory has a confined
description in terms of ``meson'' fields
$M^{jk}$ satisfying a quantum constraint
$\Pf(M) = \hbox{constant}$.
This constraint is incompatible with the condition that
$\partial W / \partial S_{jk} = 0$, and \susy is spontaneously
broken \cite{deform}.

We will write the effective lagrangian assuming that $\avg{S}$
is small enough that the confined description is valid.
We use fields $\hat{M}$ in terms of which the effective lagrangian
has the form of \Eq{shouldform}.
We have argued above that the strong dynamics gives rise to \vevs
for such fields of order 1 (in units where $\La = 1$).
Therefore, the fields $\hat{M}$ satisfy
\beq[quantumconstr]
\Pf{\hat{M}} = c \sim 1.
\eeq
The effective lagrangian written in terms of these fields is
\beq
\bal
\scr{L}_{\rm eff} &= \myint d^2\th d^2\bar{\th}\, \left[
S^\dagger S + \frac{1}{16\pi^2} \hat{K}_{\rm dyn}(\hat{M}, \la S)
\right]
\\
&\qquad
+ \myint d^2\th\, \frac{a \la}{16\pi^2} S_{jk} \hat{M}^{jk} + \hc,
\eal
\eeq
where $\hat{K}_{\rm dyn}$ is the dynamically generated
\Kahler potential.%
\footnote{$\hat{K}_{\rm dyn}$ will also contain terms depending on
\susy covariant derivatives of $\hat{M}$ and $S$.}
The coupling $a$ in the effective superpotential is an unknown
strong-interaction parameter of order 1.

To understand the physics of this lagrangian, we write
\beq
\hat{M} = \avg{\hat{M}} + \hat{M}',
\quad
S = S_0 + S',
\eeq
where $S'$ is defined by the constraint
\beq
\avg{\hat{M}^{jk}} S'_{jk} = 0.
\eeq
(There is a nonlinear constraint on $\hat{M}'$ arising
from \Eq{quantumconstr}.)
With these definitions, we see that the effective superpotential
gives $\hat{M}'$ and $S'$ a \susy-invariant Dirac mass.
Taking into account that the kinetic term for $\hat{M}'$ has a
coefficient of order $1/(16\pi^2)$, we see that this mass is of
order
\beq
m \sim \frac{\la}{4\pi} \La.
\eeq

We would like to integrate out $\hat{M}$ and $S'$ and write the
effective theory below the scale $m$.
The only light degree of freedom in this case is the field $S_0$.
It is not hard to see that the leading contributions come from
1-loop diagrams where all vertices have the form
$(S_0)^p (\hat{M}')^q$.
The resulting effective lagrangian has the form (putting back factors
of $\La$)
\beq
\bal
\scr{L}_{\rm eff} &= \myint d^2\th d^2\bar{\th}\, \left[
S_0^\dagger S^{\vphantom{\dagger}}_0
+ \left( \frac{\la \La}{16\pi^2} \right)^2
\hat{K}_{\rm dyn}(\la S_0 / \La)
\right]
\\
&\qquad
+ \myint d^2\th\, \frac{a \la \La^2}{16\pi^2} S_0 + \hc,
\eal
\eeq
In this description, the model breaks \susy due to the combination
of the linear term
in the superpotential and the non-trivial \Kahler potential.
The \susy-breaking order parameter is
\beq
\avg{F_{S_0}} \sim \frac{\la \La^2}{16\pi^2}.
\eeq
The fermionic component of $S_0$ is an exactly massless Goldstino.
The sign of the $S_0$ scalar mass term at the origin is determined
by the signs of unknown coefficients in the effective \Kahler potential,
so we cannot determine whether $\avg{S_0}$ is nonzero.
As long as $\avg{S_0}$ is small enough for the confined description
to be valid, the scalar components of $S$ have masses of order
\beq
m_0 \sim \left( \frac{\la}{4\pi} \right)^4 \La.
\eeq
The factors of $4\pi$ reduce the mass by a factor of $10^{-4}$ in
this model!

\section{Conclusions}
All of our results for strongly-interacting theories with a single 
mass scale can be summarized by the formula
\beq[final]
\scr{L}_{\rm eff} = \frac{1}{16\pi^{2}} \hat{\scr{L}}_{\rm eff},
\eeq
where all couplings in $\scr{L}_{\rm eff}$ are of
order 1 in units of the mass scale $\La$ that sets the scale for the 
``new physics'' associated with the strong dynamics.
\Eq{final} ensures that both the fundamental
and the effective theory are strongly coupled at the scale $\La$,
in the sense that all loops are equally important (``loop democracy'').
Below the scale $\La$, the effective theory may be weakly coupled,
either because the symmetries of the effective theory do not allow
renormalizable interactions (as in the case of Nambu--Goldstone bosons),
or because the renormalizable couplings that appear renormalize
logarithmically to zero in the infrared (as in confining theories).
The arguments presented here in support of this counting are at best
heuristic, but the low-energy behavior of QCD seems to support them.%
\footnote{Note that the parameter $1/(16\pi^2)$ can be viewed as being
small because the dimension of spacetime is large.
It is hard to see how to construct a systematic argument based
on this observation, since all interactions become irrelevant for large
spacetime dimension.}
As illustrated in this paper, the simple scaling embodied in 
\Eq{final} has interesting implications for dimensional analysis
in strongly-interacting \susy breaking theories.

\section{Acknowledgments}
I thank H. Murayama and S. Thomas
for discussions on the subject of this paper.
I thank R. Sundrum and J. Terning for comments on
the manuscript.


\end{document}